\documentclass[Proceedings]{SciPost2}
\usepackage{graphicx}
\usepackage{subcaption}

\usepackage{url}
\usepackage{hyperref}
\definecolor{nicered}{rgb}{0.5,0.,0.}
\definecolor{nicegreen}{rgb}{0.,0.5,0.}
\definecolor{niceblue}{rgb}{0.,0.,0.5}
\hypersetup{colorlinks,citecolor=nicegreen,linkcolor=nicered,urlcolor=niceblue}

\usepackage{amsmath,amssymb}
\usepackage{physics}
\DeclareSymbolFont{usualmathcal}{OMS}{cmsy}{m}{n}
\DeclareSymbolFontAlphabet{\mathcal}{usualmathcal}

\begin{document}


\begin{center}{\Large \textbf{
Impact of heavy-quark production measurements in the CT18 global QCD analysis of PDFs}
}\end{center}

\begin{center}
Marco Guzzi\textsuperscript{1$\star$}, Pavel Nadolsky\textsuperscript{2}, and Keping Xie\textsuperscript{3}
\end{center}

\begin{center}
{\bf 1} Department of Physics, Kennesaw State University,  Kennesaw, GA 30144, USA\\
{\bf 2} Department of Physics, Southern Methodist University, Dallas, TX 75275-0175, USA\\
{\bf 3} PITT PACC, Department of Physics and Astronomy, University of Pittsburgh, Pittsburgh, Pennsylvania 15260, USA\\
\vspace{0.5cm}
$\star$mguzzi@kennesaw.edu
\end{center}

\begin{center}
\today
\end{center}

\pagestyle{fancy}
\fancyhead[LO]{\colorbox{scipostdeepblue}{\strut \bf \color{white}~Proceedings}}
\fancyhead[RO]{\colorbox{scipostdeepblue}{\strut \bf \color{white}~DIS 2021}}


\definecolor{palegray}{gray}{0.95}
\begin{center}
\colorbox{palegray}{
  \begin{tabular}{rr}
  \begin{minipage}{0.1\textwidth}
    \includegraphics[width=22mm]{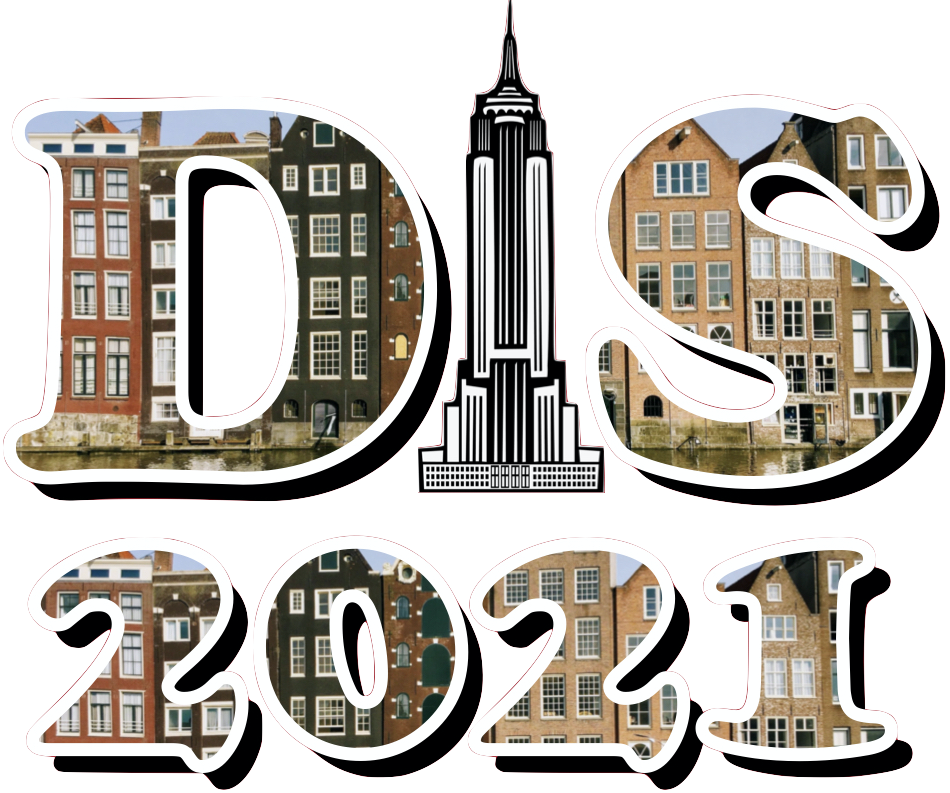}
  \end{minipage}
  &
  \begin{minipage}{0.75\textwidth}
    \begin{center}
    {\it Proceedings for the XXVIII International Workshop\\ on Deep-Inelastic Scattering and
Related Subjects,}\\
    {\it Stony Brook University, New York, USA, 12-16 April 2021} \\
    \doi{10.21468/SciPostPhysProc.xx.xxxx}
    \end{center}
  \end{minipage}
\end{tabular}
}
\end{center}

\section*{Abstract}
{\bf
We discuss the impact of heavy-flavor production measurements in semi-inclusive deep inelastic scattering at HERA on the CTEQ-TEA PDFs.
In particular, we study the impact of the latest charm and bottom production measurements from the H1 and ZEUS collaborations on the gluon,  
and the interplay of these measurements with the data ensemble of the recent CT18 global QCD analysis. 
 }

\vspace{10pt}
\noindent\rule{\textwidth}{1pt}
\tableofcontents\thispagestyle{fancy}
\noindent\rule{\textwidth}{1pt}
\vspace{10pt}

\section{Introduction}
\label{sec:intro}
Heavy-flavor production in deep inelastic scattering (DIS) reactions is important to probe 
factorization in perturbative quantum chromodynamics (pQCD) in presence of several hard scales, such as the heavy-quark masses 
and transverse momenta of the outgoing quarks. 
Various amended versions of the factorization theorem have been developed to study the production of heavy flavors in DIS, and have been 
extensively studied in literature~\cite{Aivazis:1993kh,Aivazis:1993pi,Buza:1996wv,Thorne:1997ga,Kramer:2000hn,Tung:2001mv,Forte:2010ta,Alekhin:2009ni,Guzzi:2011ew,Helenius:2018uul}.
General mass treatments are currently employed in modern global QCD analyses of parton distribution functions (PDFs) of the proton \cite{Hou:2019efy,Bailey:2020ooq,NNPDF:2017mvq,Alekhin:2017kpj}
and are critical to correctly account for phase space suppression and other mass effects that are comparable in magnitude to next-to-next-to-leading order (NNLO) 
radiative corrections in the QCD strong coupling $\alpha_s$. 
Moreover, the dependence of standard candle cross sections on heavy-quark masses $m_c$ and $m_b$, is not negligible. 

The recently published CT18 global analysis~\cite{Hou:2019efy} includes neutral current (NC) DIS measurements 
of charm and bottom structure functions $F_2^{c \bar{c}}$ and $F_2^{b \bar{b}}$ at high $Q^2$~\cite{H1:2004esl} 
as well as charm production cross section measurements at HERA~\cite{H1:2012xnw}.
These data are very important for PDF determinations, as they set direct constraints on the gluon PDF and have the potential to indirectly constrain the strange-quark PDF. 

In 2018, a new combination of charm and bottom production measurements from the H1 and ZEUS collaborations has been published~\cite{H1:2018flt} 
and superseded the previous measurements~\cite{H1:2004esl,H1:2012xnw} with an extended kinematic range 
of photon virtuality 2.5 GeV$^2$ $\leq Q^2 \leq$ 2000 GeV$^2$ and Bjorken scaling variable 
$3 \cdot 10^{-5} \leq x_\textrm{Bj} \leq 5 \cdot 10^{-2}$, 
and reduced uncertainties due to a simultaneous combination of charm and bottom cross-section measurements with reduced correlations between them.  

When these measurements replaced the previous ones in the CT18 global analysis, they could not be described with a $\chi^2/N_{pt}$ less than 1.7. 
For the CT18NNLO fit, we obtained $\chi^2/N_{pt}=1.98$ for charm production ($N_{pt}=47$), and $\chi^2/N_{pt}=1.25$ for bottom production ($N_{pt}=26$).
The CT18XNNLO fit gives $\chi^2/N_{pt}=1.71$ for charm and 1.26 for bottom production.

Tensions were observed between these new combined data and several CT18 datasets such as the LHCb 7 and 8 TeV $W/Z$ production data~\cite{LHCb:2015okr,LHCb:2015kwa}, 
$Z$-rapidity data~\cite{CDF:2010vek} at CDF run-II, CMS 8 TeV single inclusive jet production~\cite{CMS:2016lna}, and $t\bar{t}$ double differential $p_T$ and $y$ cross section~\cite{CMS:2017iqf}. 
Therefore, these data were not included in the CT18 global analysis. 

In this conference proceedings contribution, we shall come back to this point and illustrate the preliminary 
results of a more detailed analysis of these measurements~\cite{H1:2018flt}. 
We have investigated these data and explored the impact of the new 
correlated systematic uncertainties released by the H1 and ZEUS collaborations. 
The complete results of the current investigation are going to be published in a forthcoming paper~\cite{mguzziHF}.

\section{CT18 and the new charm and bottom combination at HERA}
\label{sec:challenges}

In the H1 and ZEUS combined analysis of Ref.~\cite{H1:2018flt}, heavy-quark data have been compared to the theory predictions 
obtained by different groups~\cite{H1:2015ubc,Alekhin:2009ni,Alekhin:2017kpj,Thorne:2012az,Forte:2010ta,Ball:2011mu,Ball:2017otu}
and it has been found that the $\chi^2$ for these measurements is not optimal. 
More recent global analyses~\cite{dis2021-nocera,Bailey:2020ooq} have also shown a poor description of these data. 
In all cases, the theory seems to fail to describe the slope of the data in the intermediate/small $x$ region $10^{-5}\leq x\leq 0.01$. 

In this new study, we fit the new charm and bottom combination at HERA in the CT18 data ensemble 
using the S-ACOT-$\chi$ heavy-quark scheme at NNLO~\cite{Guzzi:2011ew} which is the default 
general mass variable flavor number scheme adopted in the CTEQ global PDF analyses. 
In our attempt to fit these new measurements, we have varied several parameters in the fit 
and have explored the alternative settings in various combinations.

For example, fits with increased weights of the combined HERA data~\cite{H1:2018flt}, show preference for a harder gluon at intermediate/small $x$. 
The $\chi^2/N_{pt}$ is no less than 1.44 when the new combined HERA data are included with a large statistical weight of order 100. 
In this extreme scenario, the opposing $\chi^2$ pulls arise from the
LHCb 7 and 8 TeV W/Z cross section measurements~\cite{LHCb:2015okr,LHCb:2015kwa}, ATLAS 7 TeV~\cite{ATLAS:2016nqi}, 
CDF Run-2 inclusive jets~\cite{CDF:2008hmn}, CDF Run-2 $Z$ rapidity~\cite{CDF:2010vek}, and the D$0$ Run-2 electron $A_{ch}$ data~\cite{D0:2014kma}.

\begin{figure}[ht!]
\centering
\includegraphics[width=0.7\textwidth]{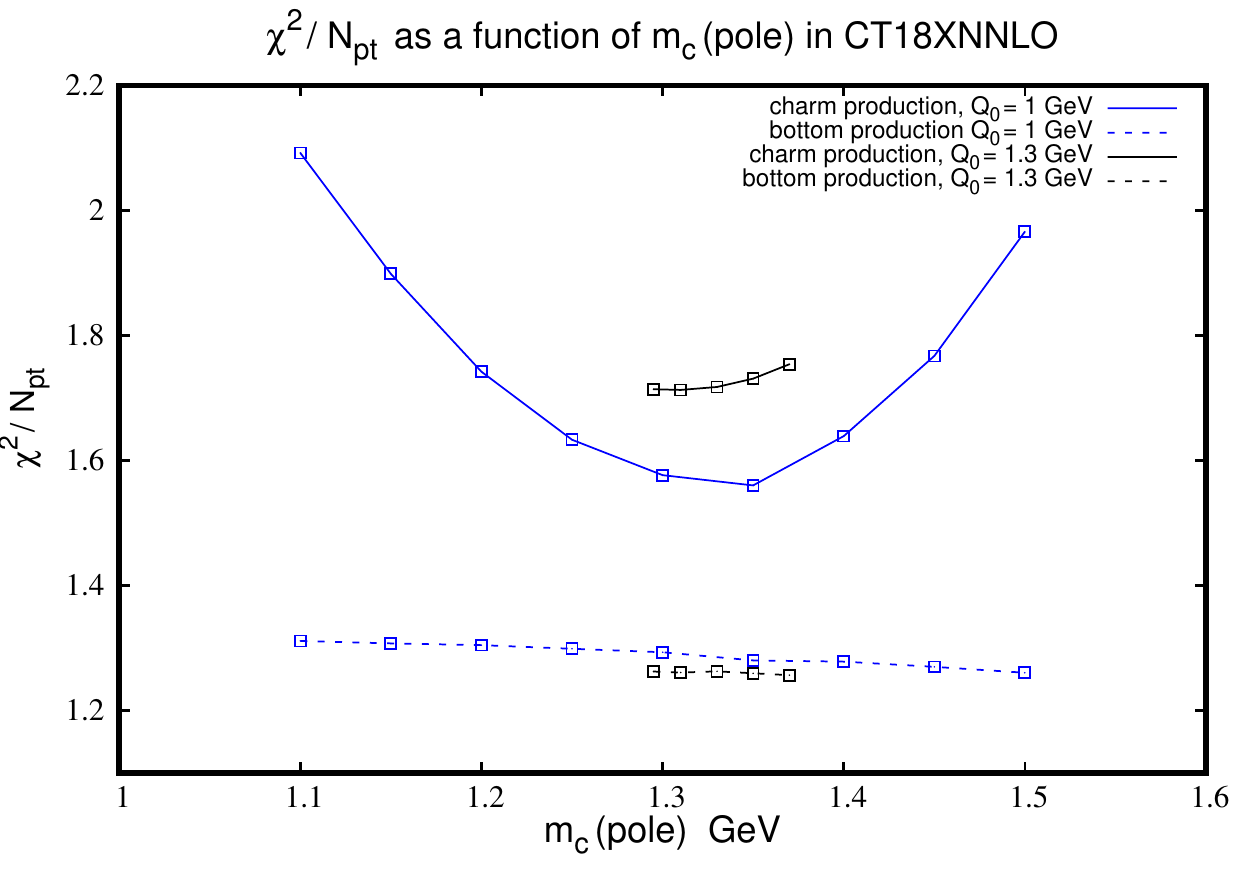}
\caption{$\chi^2/N_{pt}$ for charm (solid) and bottom (dashed) production in the CT18XNNLO fit as a function of the charm-quark mass $m_c(\textrm{pole})$ for different values of the input scale $Q_0$. }
\label{fig:chisq} 
\end{figure}

In another exercise, we varied the input charm-quark mass in either the $\overline{\textrm{MS}}$ and pole mass definitions. The initial scale $Q_0$ has also been varied consistently.   
To improve the description at intermediate/small-$x$ and deal with different initial input scale values, we tried alternative parametrizations for the gluon. 
In Fig.~\ref{fig:chisq} shows the $\chi^2/N_{pt}$ values for charm and bottom production when a scan over $m_c$(pole) is performed. 
As expected, the fit is very sensitive to the charm quark mass. However, the $\chi^2/N_{pt}$ is never lower than 1.6 in these scenarios.

\begin{figure}[h!]\hspace{-1.0cm}
\includegraphics[width=0.55\textwidth]{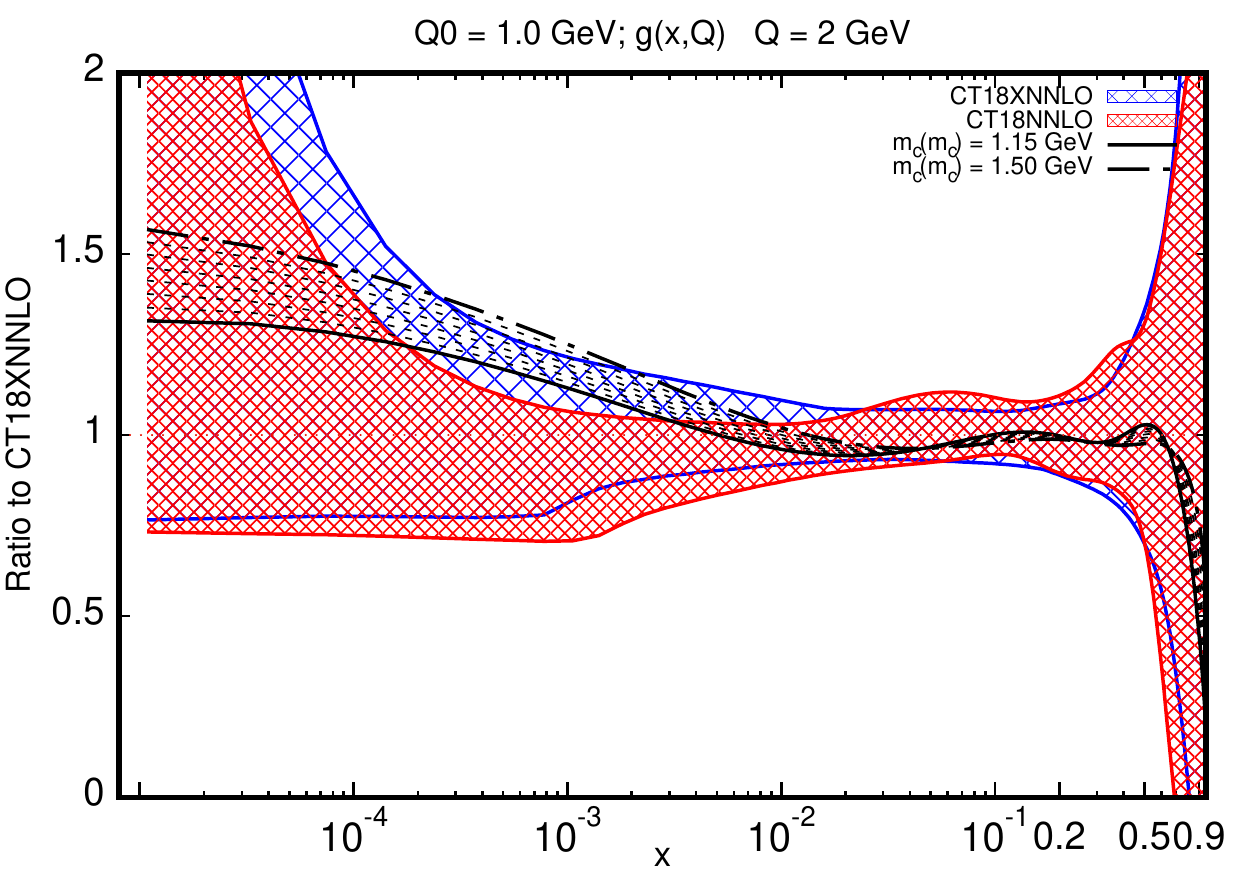}
\includegraphics[width=0.55\textwidth]{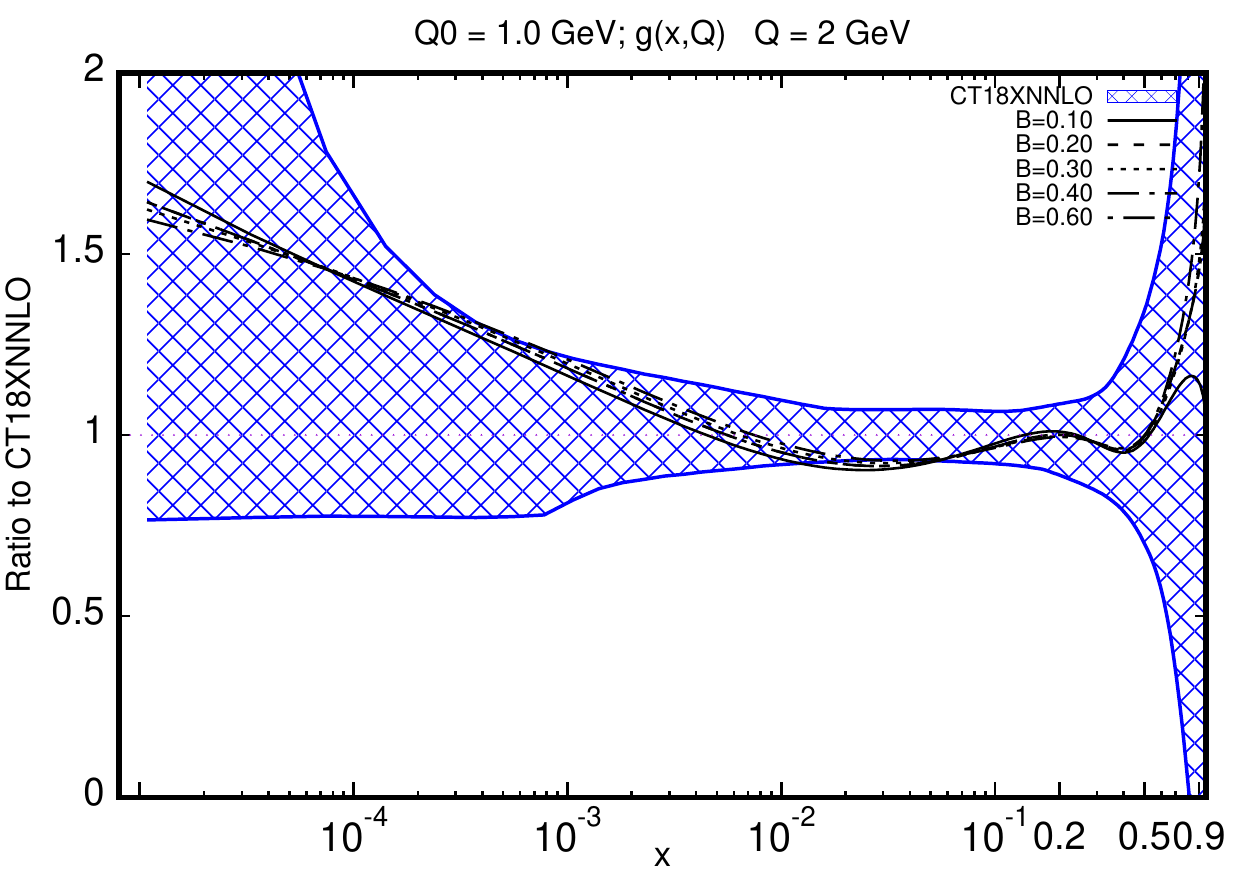}
\caption{Ratio to the CT18XNNLO gluon PDF as a function of $x$ at $Q=2$ GeV and $Q_0 = 1$ GeV. 
Left: scan over the $\overline{\textrm{MS}}$ charm-quark mass $m_c(m_c)$ while $m_b(m_b)=4.18$ GeV. Right: scan over the $B$ parameter in  $\mu_{\textrm{DIS}}(x)$. 
Error bands are shown at 90\% confidence level for CT18NNLO (red) and CT18XNNLO (blue).}
\label{fig:gluon} 
\end{figure}

We performed fits in which we varied parameters of the $x$-dependent DIS factorization scale, defined as $\mu_{\textrm{DIS}}(x) = A\sqrt{m_Q^2+B^2/x^C}$,
and used for the calculation of low-$x$ DIS cross sections in the CT18XNNLO fit. The CT18XNNLO fit is a variant of CT18NNLO   
that is generated by including the $\mu_{\textrm{DIS}}(x)$ scale choice for low-$x$ DIS data. 
This $x$-dependent scale choice mimics the main impact of low-$x$ resummation~\cite{Ball:2017otu} and is inspired by saturation models \cite{Golec-Biernat:1998zce,Caola:2009iy}.  
In fact, we observed that these data mildly prefers CT18XNNLO to CT18NNLO~\cite{Hou:2019efy}. 
In Fig.~\ref{fig:gluon}, we illustrate modifications induced on the NNLO gluon PDF at $Q=2$ GeV and $Q_0 = 1$ GeV 
when a scan over the $\overline{\textrm{MS}}$ charm-quark mass $m_c(m_c)$ is performed (left), 
and when the $B$ parameter in  $\mu_{\textrm{DIS}}(x)$ is varied. 
In the left inset, the solid black curve corresponds to the fit with $m_c(m_c)=1.15$ GeV with $\chi^2(\textrm{HERA HQ})/N_{pt}=1.62$, 
while the dotdashed represents the fit with $m_c(m_c)=1.50$ GeV and with $\chi^2(\textrm{HERA HQ})/N_{pt}=4.77$.
In the right inset, the solid black curve corresponds to the fit with $B=0.10$ GeV and $\chi^2(\textrm{HERA HQ})/N_{pt}=1.58$,
while the dotdashed represents the fit with $B=0.60$ GeV  and $\chi^2(\textrm{HERA HQ})/N_{pt}=1.52$. In both cases, parameters $A=0.5$ and $C=0.33$ are fixed in $\mu_{\textrm{DIS}}(x)$.
 
Error bands are shown at the 90\% confidence level for CT18NNLO and CT18XNNLO.   
Overall, these preliminary findings indicate that the new charm and bottom production measurements at HERA 
seem to have a preference for a harder gluon at intermediate and small $x$. 

To optimize phase-space suppression due to heavy quark masses, we have also performed fits where we varied 
the S-ACOT-$\chi$ rescaling parameter $\chi=\zeta(1+\zeta^\lambda m_Q^2/Q^2)$. It had only a modest impact on the fit.

\begin{figure}[h!] \centering
\includegraphics[width=0.64\textwidth]{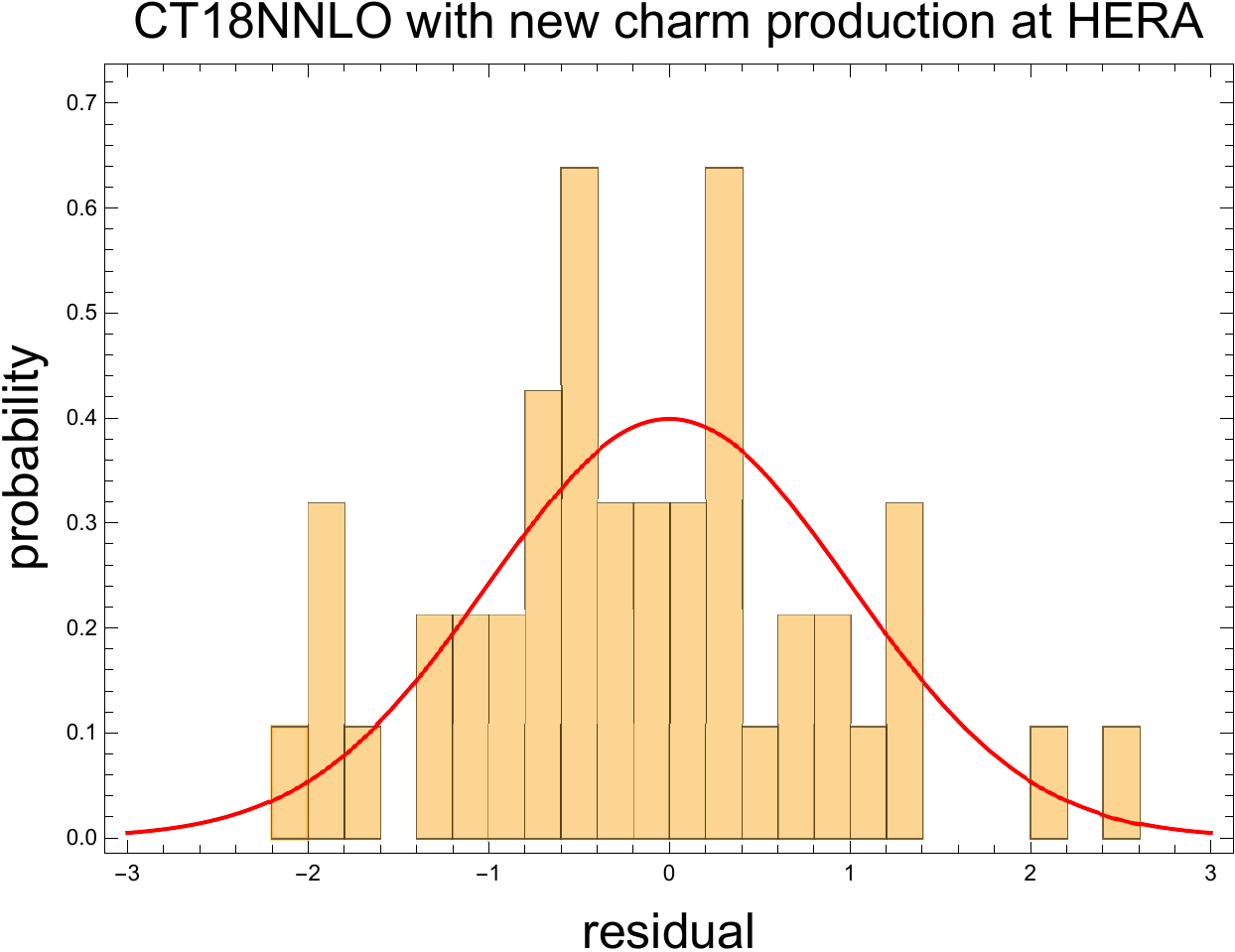}
\caption{Distribution of the residuals for the new charm production at HERA in the CT18NNLO global fit.}
\label{fig:charm-residuals} 
\end{figure}

The correlated systematic uncertainties play a very important role in the description of these data.  
In total, the H1 and ZEUS collaborations released 167 sources of correlated systematic uncertainties for the new charm and bottom combination, 71 of which 
are experimental systematic sources, 16 are related to the extrapolation procedures (i.e. fragmentation fractions and branching ratios), 
and 80 are statistical correlations between charm and bottom quarks.
In Fig.~\ref{fig:charm-residuals} we illustrate the distribution of the statistical residuals for the new charm data in the CT18NNLO global analysis, 
where the old charm and bottom production data have been replaced by the new ones. 
The residuals $r_k$ are defined in terms of the theory $T_k$ and the shifted data values $D^{sh}_k$ 
as $r_k =  (D^{sh}_k - T_k)/\sigma_k$ where $\sigma_k$ is the total uncorrelated uncertainty (more details in Ref.~\cite{Hou:2019efy}). 
The Anderson-Darling test applied to the residuals gives a $p$-value of 0.92. 
On the other hand, the distribution of nuisance parameters does not exhibit such a good behavior and the same test gives a $p$-value of $10^{-7}$.    

\section{Conclusions}
\label{sec:conclusions}
We have discussed the preliminary results of a detailed study of the new charm and bottom production combination at HERA~\cite{H1:2018flt} in the context of the CT18 NNLO global analysis.
These measurements are of high importance for PDF determinations because they provide direct constraints on the gluon PDF at intermediate and small $x$, 
and indirect constraints on strange quark PDF.
We tried to improve the description of these data within the CT18 global analysis, and performed a large number of fits   
in which we varied several parameters. These parameters are correlated to various extents 
and make their study very intricate. In the best configuration, the $\chi^2/N_{pt}$ is no lower than 1.5. 
We observed that these data seem to prefer a harder gluon in the intermediate/small $x$ region.
The $\chi^2/N_{pt}$ values which we have found are similar to those in recent studies by  
MSHT20 and NNPDF4.0, and from other groups as reported in Tab 4 of the H1 and Zeus Collaborations study~\cite{H1:2018flt}. 

A good description of these data remains challenging. A more extensive analysis of these important measurements in
the context of the CTEQ global analysis will be presented in a forthcoming study~\cite{mguzziHF}.

\section*{Acknowledgements}
The work of M.G. is supported by the National Science Foundation under Grant No. PHY-1820818. 
The work of PN is supported by the U.S. Department of Energy under Grants No. DE-SC0010129.
The work of K.X. is supported by grant No. DE-FG02-95ER40896, the National Science Foundation under grant No. PHY-1820760, and in part by the PITT PACC.
 \bibliographystyle{SciPost_bibstyle} 
\bibliographystyle{utphys}


\end{document}